\documentclass{aa}
\usepackage{upgreek}
\usepackage{bm}
\usepackage{txfonts}
\usepackage{natbib}
\usepackage{nicefrac}
\usepackage{graphicx} 
\usepackage{placeins}
\usepackage{subcaption}
\usepackage[font=small,skip=2pt]{caption}
\usepackage{hyperref}            

\hypersetup{colorlinks = {true},
            linkcolor = [rgb]{0,0.35,0.7},
            citecolor = [rgb]{0,0.35,0.7},
            filecolor = [rgb]{0.61,0,0},
            urlcolor = [rgb]{0.61,0,0},
           }

\graphicspath{ {./} }

\newcommand{\tensorGR}[1]{\overline{\bm{{#1}}}}
\newcommand{\DP}[2]{\frac{\partial{#1}}{\partial{#2}}}

\newcommand{\cs}{c_\text{s}}

\newcommand{\OmegaK}{\Omega_\text{K}}
\newcommand{\vel}{\bm{u}}
\newcommand{\nunum}{\nu_\mathrm{num}}
\newcommand{\nunumcode}{\bar{\nu}_\mathrm{num}}

\renewcommand{\eqref}[1]{Eq.~(\ref{#1})}
\newcommand{\figref}[1]{Fig.~\ref{#1}}
\newcommand{\figureref}[1]{Figure~\ref{#1}}
\newcommand{\tabref}[1]{Table~\ref{#1}}
\newcommand{\secref}[1]{Sect.~\ref{#1}}

\newcommand{\appref}[1]{Appendix~\ref{#1}}

\defcitealias{speith-2003}{SK03}

\begin{document}

\title{Measuring the numerical viscosity in simulations of protoplanetary disks in Cartesian grids}
\subtitle{The viscously spreading ring revisited}

\author{Jibin Joseph\inst{\ref{inst1},\ref{inst3}},
		Alexandros Ziampras\inst{\ref{inst1},\ref{inst2}},
    	Lucas Jordan\inst{\ref{inst1}},
    	George A. Turpin\inst{\ref{inst2}},
    	Richard P.~Nelson\inst{\ref{inst2}}}

\authorrunning{Joseph et al.}

\institute{
Institut f\"ur Astronomie und Astrophysik, Universität T\"ubingen, Auf der Morgenstelle 10, D-72076, Germany \label{inst1} 
\and Astronomy Unit, Department of Physics and Astronomy, Queen Mary University of London, London E1 4NS,UK \label{inst2} 
\and Leibniz-Institut f\"ur Astrophysik Potsdam (AIP), An der Sternwarte 16, 14482 Potsdam, Germany. \label{inst3}\\
\email{jjoseph@aip.de,lucas.jordan@uni-tuebingen.de,\{a.ziampras, g.a.turpin, r.p.nelson\}@qmul.ac.uk}}

\date{Accepted date: 24 July 2023}

\abstract
{
	Hydrodynamical simulations solve the governing equations on a discrete grid of space and time. This discretization causes numerical diffusion similar to a physical viscous diffusion, whose magnitude is often unknown or poorly constrained. With the current trend of simulating accretion disks with no or very low prescribed physical viscosity, it becomes essential to understand and quantify this inherent numerical diffusion, in the form of a numerical viscosity.}
{
	We study the behavior of the viscous spreading ring and the spiral instability that develops in it. We then use this setup to quantify the numerical viscosity in Cartesian grids and study its properties.}
{
	We simulate the viscous spreading ring and the related instability on a two-dimensional polar grid using \texttt{PLUTO} as well as \texttt{FARGO}, and ensure the convergence of our results with a resolution study. We then repeat our models on a Cartesian grid and measure the numerical viscosity by comparing results to the known analytical solution, using \texttt{PLUTO} and \texttt{Athena++}.
}
{
	We find that the numerical viscosity in a Cartesian grid scales with resolution as approximately $\nunum\propto\Delta x^2$ and is equivalent to an effective $\alpha\sim10^{-4}$ for a common numerical setup. We also show that the spiral instability manifests as a single leading spiral throughout the whole domain on polar grids. This is contrary to previous results and indicates that sufficient resolution is necessary in order to correctly resolve the instability.}
{
	Our results are relevant in the context of models where the origin should be included in the computational domain, or when polar grids cannot be used. Examples of such cases include models of disk accretion onto a central binary and inherently Cartesian codes.
}


\maketitle

\section{Introduction}\label{sec:intro}

Hydrodynamical simulations are a useful tool in studying a variety of astrophysical processes. An example of such a process is the accretion flow of a protoplanetary disk around a star. Accretion is thought to be achieved via turbulence operating in parts of the disk \citep{lyra-umurhan-2019}, that transports angular momentum radially outwards similar to a viscosity \citep{lynden1974, balbus-papaloizou-1999}, or via a magnetic torque that is applied on the material close to the surface of the disk as a stellar wind expels gas from the disk surface layers \citep{bai2013} with bulk of the disk remaining inviscid and laminar. The result, particularly in protoplanetary disks, is a steady radial infall of gas that ultimately depletes the disk over typical timescales of 1--10~Myr \citet{haisch-2001}.
	
While turbulent angular momentum transport has been the traditional way of modeling accretion, numerous recent observations of disks around T~Tauri stars \citep[e.g., the DSHARP survey,][]{andrews-etal-2018} suggest very low $\alpha\lesssim10^{-4}$ in order for turbulence to be compatible with the radial width of observed rings \citep{dullemond-etal-2018}, the vertical structure of mm grains \citep{dullemond-etal-2022} and the formation of rings and gaps by embedded planets \citep{zhang-etal-2018}. As a result, magnetic winds are becoming a favored means of interpreting accretion in protoplanetary disks, and numerical models of such disks tend towards the inviscid limit \citep[e.g.,][]{lega-etal-2022}. 

Here, however, we run into a different problem: the numerical schemes used to model protoplanetary disks introduce a certain amount of numerical diffusion. The magnitude of this diffusion, is often unknown or poorly constrained. Thus, calculating an upper limit to this non-physical diffusion, or equivalently a "numerical viscosity", is essential in ensuring that results are not affected by the effects it can induce. This is particularly important for models using Cartesian grids, which are primarily used when the central object should be included in the simulation domain such as when modeling accretion patterns around binary stars \citep[e.g.][]{tiede-2021}, or for certain MHD codes \citep[e.g.,][]{fromang-etal-2006}. Analogue issues also appear when an object is not centered on a polar grid, which is often the case for circumplanetary disks \citep[e.g.,][]{crida2009}. 
Grid-noise due to the asymmetric nature of Voronoi-mesh cells has also been identified as a source of numerical diffusion in moving mesh codes \citep{zieler2022}. In these cases, the geometry of the grid introduces a very high numerical viscosity and as a result models require the execution of very computationally expensive, high-resolution simulations.

In this work, we revisited the viscously spreading ring problem introduced first by \citet{luest1952} and again by \citet{lynden1974}, and analyzed in 2D by \citet{speith-2003} (henceforth \citetalias{speith-2003}). We approached this with numerical hydrodynamics simulations that first aimed to reproduce or improve upon the results of \citetalias{speith-2003} by reanalyzing the azimuthal instability they reported, and then performed calculations in Cartesian coordinates to quantify the numerical viscosity of the codes \texttt{PLUTO} and \texttt{Athena++}.

The viscous ring problem and our physical and numerical setup are described in Sect.~\ref{sec:model}. We compare our results with those of \citetalias{speith-2003} and point out the origin of a numerical artifact in their study through a resolution analysis in Sect.~\ref{sec:standard_model}. We present our results for Cartesian models and our estimates of the numerical viscosity in Sect.~\ref{sec:viscous-ring-cartesian}. We then discuss our findings in Sect.~\ref{sec:discussion}, and conclude in Sect.~\ref{sec:conclusions}.

\section{Model setup}\label{sec:model}
In this section, we revisit the setup for the viscously spreading ring problem and describe our physics and numerical methods. We also provide a list of models that were executed using different codes and in different coordinate systems.
\subsection{The viscous ring problem} 
Starting from an axisymmetric surface density distribution of a pressureless ($P=0$) ring of gas with mass $M_\text{ring}$ that is centered at distance $R_0$ and is infinitely thin in the radial direction
\begin{equation}
	\Sigma_\mathrm{ring}(t=0, R) = \frac{M_\mathrm{ring}}{2\pi R} \delta (R - R_0),
\end{equation}
the ring will spread along the radial direction $R$ following the analytical solution \citep[e.g.,][]{lynden1974,pringle-1981}

\begin{equation}
\label{eq:viscous-ring}
\Sigma_\mathrm{ring}(\tau, x)  = \frac{M_\mathrm{ring}}{\pi R_0^2} \frac{1}{\tau 
x^{1/4}} \, I_\frac{1}{4}\left(\frac{2x}{\tau}\right) \mathrm{exp}\left( - 
\frac{1 + x^2}{\tau} \right),
\end{equation}
where $x = R/R_0$ and $\tau = 12\nu tR_0^{-2}$ are normalized distance and time quantities, $\nu$ is the kinematic viscosity of the gas, and $I_\frac{1}{4}$ is the modified Bessel function of the order $1/4$. In addition, the radial velocity of the ring is given by the following relation,
\begin{equation}
\label{eq:vel-radial}
u_\mathrm{R,ring} = -\frac{3}{\Sigma_\mathrm{ring}\,
\sqrt{R}}\frac{\partial}{\partial R} \left[\nu \Sigma_\mathrm{ring} 
\sqrt{R}\right]
\end{equation}
We note that, \eqref{eq:viscous-ring} is only an approximate solution to the hydrodynamic equations (see \secref{sec:hydro}), and assumes that the disk is massless with highly supersonic and Keplerian azimuthal velocity.  Furthermore, one assumes that the kinematic viscosity, $\nu$, is much smaller than the specific angular momentum $R^2\Omega$, an assumption that is violated close to the central object. A time evolution of the viscous ring problem using our fiducial axisymmetric numerical setup (described in Sect.~\ref{sub:numericalsetup}) is shown in Fig.~\ref{fig:1Dpol-std-evo}.

The problem was revisited by \citetalias{speith-2003}, who showed that the ring is subject to a viscosity-driven spiral instability in the azimuthal direction that results in the development of a leading spiral arm that covers the full extent of the computational domain (see Fig.~\ref{fig:2Dpol-std-evo}). They performed a detailed stability analysis and showed that unstable modes with wave-numbers $k$ develop where
\begin{equation}
	\label{eq:stability-criterion}
	k^2 > \frac{3}{R \Sigma_0} \frac{\partial \Sigma_0}{\partial R}
\end{equation}
with $\Sigma_0$ being the axisymmetric density profile. They further showed that the spiral arm changes direction from leading to trailing at around the peak of the ring (see second and third panels of Fig.~\ref{fig:willy_compare_sample}), and verified their results with two fundamentally different codes: the finite-difference upwind grid code \texttt{RH2D} \citep{kley-1989,kley-1999}, and the smooth particle hydrodynamics (SPH) code used by \citet{flebbe-1994}.

\subsection{Hydrodynamics}
\label{sec:hydro}

We consider the vertically-integrated hydrodynamics equations for a gas with surface density $\Sigma$, velocity vector $\vel$, and nearly zero pressure $P$ at distance $R$,
\begin{equation}
	\label{eq:hydro}
	\begin{split}
		&\DP{\Sigma}{t} + \nabla\cdot(\Sigma\vel) = 0, \\
		&\DP{(\Sigma\vel)}{t} + \nabla\cdot\left(\Sigma\vel\otimes\vel\right) = -\nabla P - \Sigma\nabla\Phi_\star + \nabla\cdot\tensorGR{\upsigma}.
	\end{split}
\end{equation}
The gas orbits around a star with mass $M_\star$ such that the gravitational potential is given by $\Phi_\star=-\text{G}M_\star/R$, with $\text{G}$ being the gravitational constant. The viscous stress tensor is represented with $\tensorGR{\upsigma}$. We use a locally isothermal equation of state, such that the gas has a (very small) sound speed $c_\text{s}=\sqrt{P/\Sigma}=h \sqrt{\text{G}M_\star/R} = h R\OmegaK$, where $h$ is the aspect ratio.

\subsection{Numerical setup}
\label{sub:numericalsetup}

For our standard setup, we used \texttt{PLUTO} version 4.3 \citep{mignone-2007}, a finite-volume code with a second-order accurate scheme \citep[HLLC,][]{toro-etal-1994} and the van~Leer flux limiter \citep{vanleer-1977}. We followed the setup laid out in \citetalias{speith-2003} to reproduce the original results. In this section, we provide a detailed description of our model and its differences with the setup in \citetalias{speith-2003}.

To study the viscous ring and the related instability we used a polar grid with computational domain $R \in [0.2,2]\,R_0$ and $\phi \in [0,2\pi)$. We used both logarithmic ($\Delta R \propto R$) and arithmetic ($\Delta R=\mathrm{const.}$) scaling in the radial direction for comparison reasons, but focus on logarithmic setups first. We also enabled the \texttt{FARGO} transport scheme \citep{masset-2000} implemented in \texttt{PLUTO} by \citet{mignone-etal-2012}, while the parabolic viscosity term is handled using the explicit time stepping method in \texttt{PLUTO}. We used a Courant number of 0.4 in all simulations, unless specified otherwise. The system was then evolved for various grid resolutions, the results of which are discussed in Sect.~\ref{sub:resolution-study}. The Cartesian setup used to measure the numerical viscosity is described in \secref{sub:numerical-viscosity}.

To evolve the viscous ring problem, we initialized the surface density using \eqref{eq:viscous-ring} at $\tau_0=0.018$ and added a small, constant surface density floor $\Sigma_\mathrm{floor} = 10^{-7}\,\Sigma_\mathrm{ref}$, where $\Sigma_\mathrm{ref} = \Sigma_\mathrm{ring}(\tau_0, 1)$. We chose a set of code units where $\bar{\mathrm{G}} = \bar{M}_\star = \bar{R}_0 = 1$. This leaves the kinematic viscosity as the only relevant physical constant in this setup, which we chose to be $\bar{\nu} = 10^{-5}$. For this viscosity, we get a viscous ring spreading time (i.e., $\tau = 1$) of $1326$ orbital periods ($P_0$) at $R = R_0$. To facilitate development of the spiral instability, the initial surface density distribution was seeded with a small amount of noise and evolved up to $t=400~P_0$, or $\tau = 0.3$.

We chose a strict outflow boundary condition in both radial directions such that $\partial_R\Sigma=0$, $u_{R,\text{in}}=-|u_R(R_\text{in})|$, $u_{R,\text{out}}=|u_R(R_\text{out})|$. The azimuthal velocity is fixed to the Keplerian speed corrected for pressure support $u_\phi=R\OmegaK\sqrt{1-h^2}$ at the boundaries. 
We explored various values of the aspect ratio, shown in \appref{apdx:aspect_ratio}, and chose $h = 0.005$ for our simulations, as we found it to be low enough for the gas to act like a pressureless fluid and high enough to ensure numerical stability in our simulations. The \texttt{RH2D} setup in \citetalias{speith-2003} used a viscosity value of $\bar{\nu} = 4.77 \times 10^{-5}$ and an initial viscous time $\tau_0 = 0.016$, but this has no effect on the subsequent evolution.

To corroborate our results, we used the \texttt{FARGO} code (legacy version) \citep{masset-2000}. Similar to the \texttt{RH2D} code used in \citetalias{speith-2003}, \texttt{FARGO} uses a finite difference scheme as described in \citet{stone-1992} with the second-order upwind algorithm by \citet{vanleer-1977}. Both codes therefore have similar behavior and we used \texttt{FARGO} to compare our results to \citetalias{speith-2003}.
\section{Analysis of the viscous ring problem}
\label{sec:standard_model}
In this section, we explore the spiral instability using logarithmically spaced polar grids. In such grids, the radial cell size $\Delta R$ increases with distance such that the ratio $\Delta R/H$ remains comparable throughout the domain (in the case of $h=\mathrm{const.}$, this ratio is constant as $H=hR$). This makes logarithmically spaced grids well-suited for general-purpose disk models, where gas dynamics should be resolved over a scale height $H$ using a reasonable number of cells. We briefly describe the general behavior of the viscous spreading ring and show that our fiducial model is numerically converged.
\subsection{Time evolution of the viscous ring}
\label{sub:time-evo}
\begin{figure}[h]
	\includegraphics[width=\columnwidth]{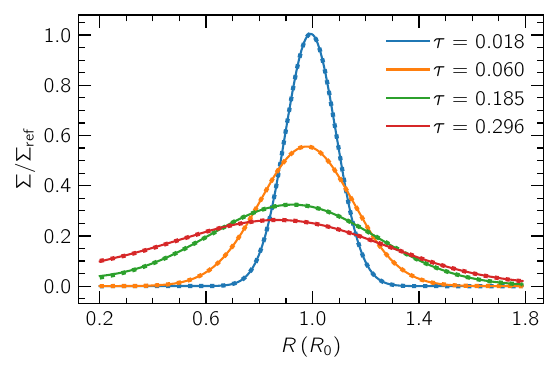}
	\caption{Evolution of the one-dimensional viscously spreading ring. The colors indicate different timestamps, solid lines represent simulation results and	dots follow the analytical solution at the corresponding time according to \eqref{eq:viscous-ring}.}
	\label{fig:1Dpol-std-evo}
\end{figure}

\begin{figure}[h]
	\includegraphics[width=\columnwidth]{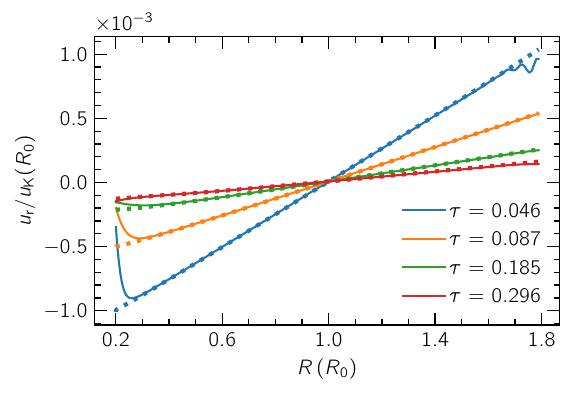}
	\caption{Similar to \figref{fig:1Dpol-std-evo} showing the evolution of radial velocity of the ring compared to the analytical solution \eqref{eq:vel-radial}.}
	\label{fig:1Dpol-vradial}
\end{figure}

We begin our analysis with a 1D model of the viscously spreading ring, using 465 cells logarithmic scaled in the radial direction. This resolution is chosen such that the gas is resolved with at least one cell per scale height $H=h R$. Even though the problem is formally defined for a pressureless fluid ($h=0$), \figref{fig:1Dpol-std-evo} and \figref{fig:1Dpol-vradial} show that our model adequately reproduces the analytical equations of \eqref{eq:viscous-ring} and \eqref{eq:vel-radial}, respectively.

To assess the robustness of the numerical methods used, we quantified numerical errors in our results by measuring the average density-weighted relative deviation from the analytical solution of both the surface density $\Sigma$ and radial velocity $u_R$ on the grid, at various times. We find a maximum deviation of $0.74\%$ for $\Sigma$ and $3.31\%$ for $u_R$. While these values are largely influenced by regions close to the boundaries, where deviations are the largest, for most of the domain (0.4--1.8\,$R_0$) we find maximum deviations of $0.15\%$ and $0.75\%$, for $\Sigma$ and $u_R$ respectively. In addition, we find that further reducing the aspect ratio $h$ does not substantially affect the measured errors.

\begin{figure*}[t]
	\includegraphics[width=\textwidth]{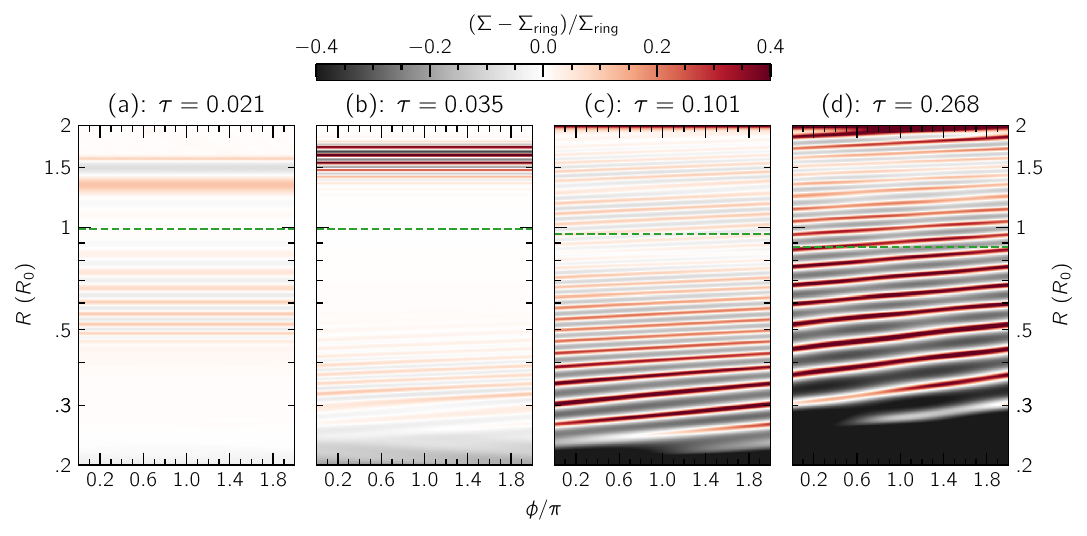}
	\caption{Snapshots of the normalized deviation from the analytical surface density distribution for our fiducial model at different times. The dashed green line marks the position of maximal surface density that indicates the peak of the spreading ring. Initial waves (a) are seen as outward-moving rings while the spiral instability develops as a spiral arm at the inner boundary (b) and spreads through the whole domain (c,d).}
	\label{fig:2Dpol-std-evo}
\end{figure*}

We then expanded to a 2D $\{R, \phi\}$ domain, with a fiducial resolution of $N_R\times N_\phi=465\times256$. Here, we expect the same radial spreading of the ring but also development of the viscosity-driven spiral instability discussed in \citetalias{speith-2003}. \figureref{fig:2Dpol-std-evo} shows a time evolution of the ring spreading and of the instability which manifests in the form of an outward-propagating spiral arm. Weak radial wave-like perturbations are launched immediately after the start of the simulation (see panel (a)). 

The perturbations are not affected by changes to pressure or initial conditions, but they do become weaker and move to smaller wavelengths with lower viscosity. This viscosity dependency is in agreement with the viscous overstability \citep[e.g.,][]{latter-2006} but a more thorough study of the perturbations is not within the scope of this work.

These initial waves leave the domain through the outer boundary while the spiral instability begins to develop near the inner boundary, with the spiral becoming apparent at $\tau\approx0.03$ (panel (b)). The spiral continues to grow in amplitude and propagate outwards until it eventually spans the entire radial extent of the disk (panels (c), (d)).

In contrast to the results in \citetalias{speith-2003}
we do not see a flip of the direction of the spiral from leading to trailing
around the peak of the ring. Instead, we find that this flip is merely a
numerical artifact and does not appear in simulations with a high enough radial resolution. We discuss this further in \appref{sec:spiral_flip}.
\begin{figure*}[t]
	\includegraphics[width=\textwidth]{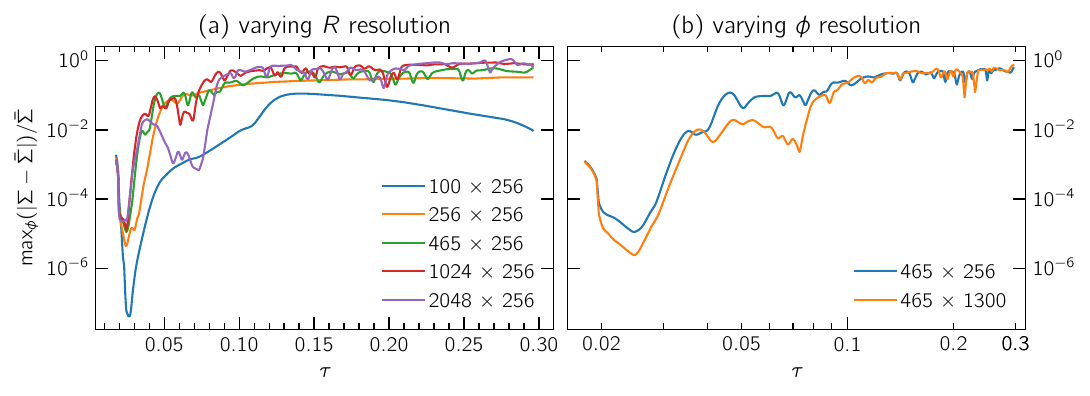}
	\caption{(a): Time evolution of the normalized maximal deviation from the azimuthally averaged surface density distribution for various numerical configurations. Inadequate resolution in the $R$-direction can damp the viscous instability, with convergence being achieved for our fiducial model with $N_R = 465$. (b): Comparison of the same quantity for different azimuthal resolutions and $N_R = 465$. The higher resolution of $N_\phi=1300$ azimuthal cells results in square grid cells but does not influence the growth of the instability.}
	\label{fig:growth-rates}
\end{figure*}
\subsection{Resolution study} \label{sub:resolution-study}
To ensure that we have properly resolved and developed the viscous instability, we performed a resolution study by analyzing its growth phase. For this, we measured the largest density deviation from the azimuthally averaged profile at different radii and tracked this quantity over time. In that sense, this quantity acts as a proxy for the time evolution of the amplitude of the strongest spiral at a given radius, which hints at the growth rate of the spiral instability. \figureref{fig:growth-rates} shows this evolution for various resolutions. Alternatively, we also evaluated the amplitude of the azimuthal Fourier modes of density rings at specific radii over time. This method yields similar results to those obtained from measuring the density deviations and are therefore not shown here.

After comparing our results against a model with square cells ($N_R\times N_\phi=465\times1300$), we found that increasing the azimuthal resolution beyond 256 cells has no effect on the growth rate of the instability (see panel (b) in \figref{fig:growth-rates}) and therefore fixed the number of azimuthal cells to 256 in our resolution study. In the radial direction we found a resolution of 465 cells to be sufficient.

We confirmed our results using the \texttt{FARGO} code in \appref{sec:spiral_flip} and achieved numerical convergence with both codes at a resolution of $465\times256$ cells. We conclude that this resolution is sufficient to fully resolve the viscously spreading ring and the development and saturation of spiral arms due to the spiral instability.
\FloatBarrier
\section{The viscous ring problem in Cartesian coordinates}
\label{sec:viscous-ring-cartesian}
Having profiled the viscous ring problem in a polar grid, we now switch to Cartesian coordinates with our ultimate goal being the measurement of the numerical viscosity of \texttt{PLUTO} for such a grid. For this set of models our computational domain extends between $x,y\in[-2,2]\,R_0$, with a fiducial resolution of $N_x\times N_y=1024\times1024$ cells which corresponds to approximately one cell per scale height at $R=\sqrt{x^2+y^2}=R_0$. We note that, $\Delta x=\Delta y=\mathrm{const.}$ in these models, and therefore our effective resolution is significantly lower at smaller radii.

For a fair comparison with our models on a polar grid, we retained the second-order accurate HLLC scheme with a linear spatial reconstruction and van Leer flux limiter for all simulations unless stated otherwise. In addition, we emulated the inner radial boundary of a polar grid using a damping region, where we damped $\Sigma$ to $\Sigma_\mathrm{ring}$ through \eqref{eq:viscous-ring} and $u_\mathrm{R}$ to zero for $R<0.2\,R_0$. Finally, we set an outflow boundary condition at all domain boundaries.
To measure numerical viscosity in \texttt{PLUTO} we used a similar setup but damped $\Sigma$ to $\Sigma_\mathrm{floor}$ following \citep{devalborro-etal-2006}, and provide a comparison using \texttt{Athena++} version 21.0 \citep{stone-etal-2008}.

\subsection{Viscous evolution}
\label{sub:viscous-cartesian}

\begin{figure}[h]
	\includegraphics[width=\columnwidth]{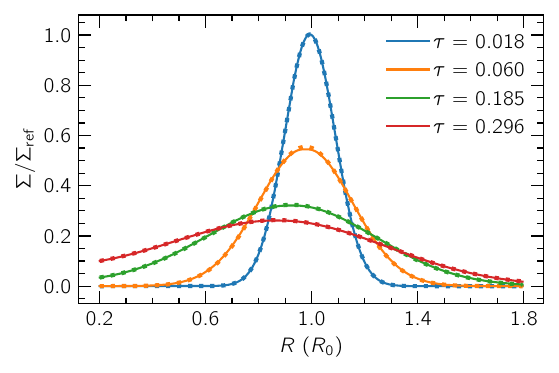}
	\caption{Snapshots of slices through the $y=0$ plane showing the evolution of surface density for our fiducial viscous Cartesian  simulation. The colors indicate different timestamps, solid lines represent simulation results and dots follow the analytical solution at the corresponding time according to \eqref{eq:viscous-ring}. Unlike the 2D polar case, no spirals develop.}
	\label{fig:1Dcart-1024-evolution}
\end{figure}

\begin{figure}[h]
	\includegraphics[width=\columnwidth]{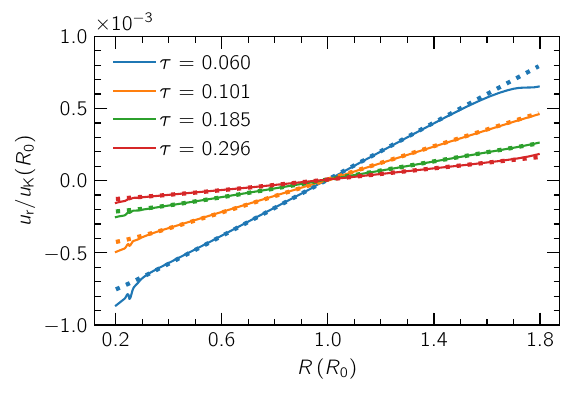}
	\caption{Similar to \figref{fig:1Dcart-1024-evolution} showing the evolution of radial velocity of the ring compared to the analytical solution \eqref{eq:vel-radial}.}
	\label{fig:1Dcart-1024-evo-vel}
\end{figure}

We first analyze the evolution of the ring with standard viscosity similar to Sec.~\ref{sec:standard_model}. The radial ring spreading and the radial velocity evolution can be seen in \figref{fig:1Dcart-1024-evolution} and \figref{fig:1Dcart-1024-evo-vel} respectively. They match the analytical solution well except for noticeable deviations in $u_\mathrm{R}$ close to the damping region and boundaries. We again quantified the error by measuring the density-weighted relative deviations in the domain excluding the damping region and found a maximum deviation of $0.71\%$ for $\Sigma$ and of $15.74\%$ for $u_\mathrm{R,ring}$. We note that, unlike the polar case, we find no development of spirals in any Cartesian model. We suspect that the spiral instability is suppressed as a result of the highly diffusive grid, further discussed in Sec.~\ref{sec:discussion}. We see similar results at a lower and higher resolution of $512^2$ and $2048^2$ cells, respectively.


\subsection{Inviscid models: estimates of numerical viscosity}
\label{sub:numerical-viscosity}

The discretization in space and time inherent in the numerical schemes employed by our codes results in numerical errors when solving the hydrodynamics equations. In our analysis in \appref{apdx:num-visc}, we show that this error has the form of a numerical viscosity and can lead to ring spreading, even in the absence of any physical viscosity. In this section, we repeat the models shown in \secref{sub:viscous-cartesian} with an inviscid prescription ($\nu=0$), and analyze the subsequent ring spreading in an attempt to quantify the numerical viscosity of the methods in \texttt{PLUTO}.

\begin{figure}[h]
	\includegraphics[width=\columnwidth]{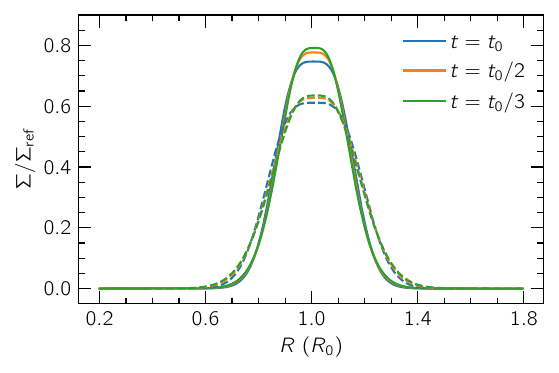}
	\caption{Slices of surface density profiles for our fiducial inviscid
		Cartesian model ($\nu=0$, blue lines) at a given time $t_0$ against the corresponding 
		viscous models with $\nu=\nunum$ at $t_0/2$ (orange) and $\nu=2\,\nunum$ at $t_0/3$ (green). Solid and dashed lines correspond to $t_0 = 224$ and 748 orbits at $R_0$, respectively. The model with $\nu=\nunum$ evolves twice as fast as the inviscid simulation and the $\nu=2\,\nunum$ model evolves thrice as fast, indicating a viscous contribution by the non-negligible numerical viscosity.}
	\label{fig:1D-cart1024-v0-comp}
\end{figure}

Blue lines in \figref{fig:1D-cart1024-v0-comp} show the time evolution for our fiducial inviscid model at a resolution of $1024\times1024$. We found that the ring spreading is indeed quite similar to viscous models, with the exception that the peak of the ring is flatter compared to the analytical solution. We then extracted a global estimate of the numerical viscosity $\nu_\mathrm{num}$ by fitting the analytical solution \eqref{eq:viscous-ring} to our data. The fitting method is described in \appref{apdx:measuring-num-visc}.

We repeated the simulation with a prescribed kinematic viscosity equal to the measured numerical viscosity ($\nu = \nunum$) and once again with $\nu = 2\,\nunum$ , the time evolution of which is shown as orange and green lines respectively in \figref{fig:1D-cart1024-v0-comp}. We found that the $\nu = \nunum$ model evolves twice as fast as our inviscid model, and the $\nu = 2\,\nunum$ model evolves thrice as fast. This suggests that all models evolve as if the total viscosity were  $\nu_\text{tot}=\nu+\nu_\mathrm{num}$. We motivate our approach and rationalize this result in \appref{apdx:num-visc}.
\subsubsection{Resolution scaling of the numerical viscosity}
\label{subsub:resolution_scaling}
\begin{figure}[h]
	\includegraphics[width=\columnwidth]{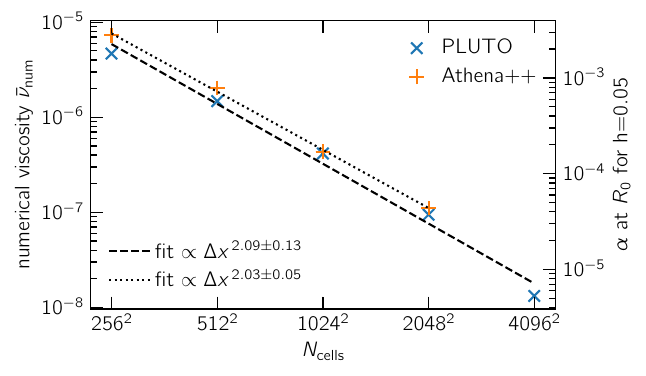}
	\caption{Numerical viscosity $\nunumcode$ in code units and the corresponding $\alpha$ at $R_0$ for $h=0.05$ at different resolutions, extracted by fitting \eqref{eq:viscous-ring} to our Cartesian grid results for both \texttt{Athena++} and \texttt{PLUTO}. The two codes agree very well for $N_\textrm{cells}\geq1024^2$. See \appref{apdx:athena} for details on the \texttt{Athena++} models.}
	\label{fig:numerical-viscosity}
\end{figure}

In our analysis of the numerical viscosity for a first-order solver in \appref{apdx:num-visc}, we show that the numerical viscosity scales as $\nu_\mathrm{num} \propto \Delta x$ for a first-order method. As we used a second-order solver for our tests, we expect the numerical viscosity to scale as $\nu_\mathrm{num} \propto \Delta x^2$ in our models. To test this, we conducted simulations for grid resolutions of $256^2$, $512^2$, $1024^2$, $2048^2$ and $4096^2$ cells. The resulting values of $\nunum$ as a function of cell count are listed in \tabref{tab:numerical_viscosities} and shown in \figref{fig:numerical-viscosity}. We then fit a power law to the relation $\nunum(\Delta x)$ and found that the numerical viscosity scales as $\nunum \propto \Delta x^{2.09}$. We also found that reducing the maximum Courant number $C_\mathrm{max} \approx \Delta t/\min\left(\nicefrac{\Delta x}{u}\right)$ from our nominal value of 0.4, had no effect on the numerical viscosity, indicating that the dependence of $\nu_\mathrm{num}$ on $\Delta x$, $\Delta t$ and the Courant number is more complicated than it would be for a first-order scheme. While we find our $\nu_\mathrm{num} \propto \Delta x^2$ estimate to roughly describe the convergence, we also find that the convergence rate is not constant and increases with resolution. We found similar results using the code \texttt{Athena++} (see Appendix~\ref{apdx:athena}).

\begin{table*}
	\centering
	\caption{Numerical viscosity $\nunum$ with one standard deviation errors in code units for different grid types and resolutions. An equivalent $\alpha_\text{num}$ at $R_0$ and a realistic $h=0.05$ is also listed.}
	\renewcommand{\arraystretch}{1.2}
	\begin{tabular}{lllll}
		\hline
		\hline
		Code & Grid & Resolution & $\nunumcode$ & $\alpha_\text{num}$ at $R_0$\\
		\hline
		\texttt{PLUTO} & Cartesian  & $256\times256$   & $(4.68\pm0.19)  
		\times10^{-6}$ & 
		$(1.87\pm0.08) \times10^{-3}$ \\
		\texttt{PLUTO} & Cartesian  & $512\times512$   & $(1.48\pm0.01) 
		\times10^{-6}$ & 
		$(5.92\pm0.04)\times10^{-4}$ \\
		\texttt{PLUTO} & Cartesian  & $1024\times1024$ & $(4.16\pm0.03) 
		\times10^{-7}$ & 
		$(1.67\pm0.01) \times10^{-4}$ \\
		\texttt{PLUTO} & Cartesian  & $2048\times2048$ & $(9.45\pm0.11) 
		\times10^{-8}$ & 
		$(3.78\pm0.04)\times10^{-5}$ \\
		\texttt{PLUTO} & Cartesian  & $4096\times4096$ & $(1.32\pm0.08) 
		\times10^{-8}$ & 
		$(5.29\pm0.31) \times10^{-6}$ \\
		\texttt{PLUTO} & polar      & $465\times256$   & $(3.89\pm0.94) 
		\times10^{-15}$ & 
		$(1.56\pm0.38) \times10^{-12}$ \\[1ex]
		\texttt{Athena++} & Cartesian  & $256\times256$   & $(7.31\pm0.22) 
		\times10^{-6}$ 	& $(2.92\pm0.08) \times10^{-3}$ \\
		\texttt{Athena++} & Cartesian  & $512\times512$   & $(2.04\pm0.05) 
		\times10^{-6}$ & $(8.16\pm0.19) \times10^{-4}$ \\
		\texttt{Athena++} & Cartesian  & $1024\times1024$ & $(4.26\pm0.11) 
		\times10^{-7}$ & $(1.70\pm0.04) \times10^{-4}$ \\
		\texttt{Athena++} & Cartesian  & $2048\times2048$ & $(1.11\pm0.02) 
		\times10^{-7}$ & $(4.44\pm0.08) \times10^{-5}$ \\
		\hline
		\label{tab:numerical_viscosities}
	\end{tabular}

\end{table*}
For a realistic protoplanetary disk with $h\sim0.05$, the numerical viscosity $\nunumcode = 4.16 \times 10^{-7} \sqrt{\mathrm{G}M_0 R_0}$ extracted from our fiducial Cartesian model translates to an $\alpha$ viscosity parameter  \citep{shakura-sunyaev-1973} of $\alpha \sim 1.67\times10^{-4}/\sqrt{\bar{R}}$. More generally, we can write
\begin{equation}
	\alpha = \frac{\nu}{\cs H} \approx 400\times \bar{\nu}\, \left(\frac{0.05}{h}\right)^2\,\sqrt{\frac{R_0}{R}},
\end{equation}

\section{Discussion}
\label{sec:discussion}

In this section, we compare our results on the spiral instability to those found in  \citetalias{speith-2003}, and comment on the nature of numerical viscosity.

\subsection{Comparison to previous results}
Analyzing the spreading ring on a polar grid in \secref{sec:standard_model}, we could confirm the main findings from \citetalias{speith-2003} that the viscous pressureless spreading ring is subject to a viscous instability that manifests as a leading spiral arm. On the other hand, we found their described flip from a leading to a trailing spiral to be purely of numerical origin. As shown in \appref{sec:spiral_flip}, we attribute the spiral flip to an under-resolved inner disk; the quantity $\Delta R/R$ increases for small radii in an arithmetic grid like that of \citetalias{speith-2003}, while a constant smoothing length in their SPH simulations results in effectively the same issue. In our adequately resolved models ($N_R\ge465$), we find many azimuthal spiral modes growing at the beginning of our simulation with their amplitude decaying exponentially with mode number. In our arithmetic, low-resolution analog of \citetalias{speith-2003}, however, the growth of azimuthal Fourier mode amplitudes is delayed, weaker and their amplitude decays faster with increasing mode number (\figref{fig:fargo_comparewillygrowth}). We provide more details in \appref{sec:spiral_flip}, but conclude here that any effects beside the leading spiral found in \citetalias{speith-2003} was due to their poor numerical resolution.
\subsection{Differences between physical and numerical viscosity}
\begin{figure}[h]
	\includegraphics[width=\columnwidth]{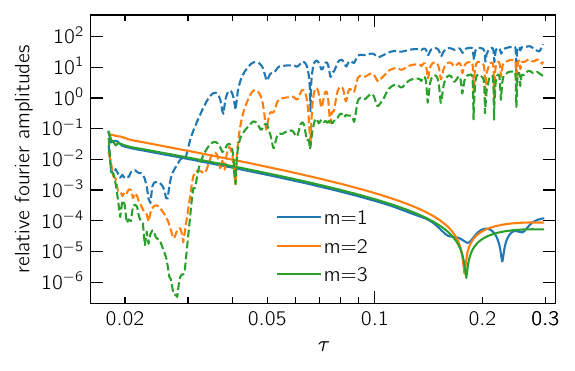}
	\caption{Fourier amplitude normalized to azimuthally averaged density at $R = 0.5\,R_0$ for $m = 1,2,3$ for the fiducial polar simulation (dashed lines) and interpolated fiducial Cartesian simulation (solid lines). The spiral instability does not develop, with amplitudes up to six orders of magnitude weaker compared to the polar case.}
	\label{fig:fid-polar-cart-fft-comp}
\end{figure}

In \secref{sub:time-evo} we showed the development of the spiral instability for our fiducial polar model (\figref{fig:2Dpol-std-evo}) and its growth (\figref{fig:growth-rates}). Unlike the polar grid, our Cartesian models did not develop this instability, even at comparatively higher resolutions. This can be attributed to the lack of angular momentum conservation, which results in a dramatically high numerical diffusivity. We confirm this by analyzing the growth rates of the instability just like for the polar case in \figref{fig:fid-polar-cart-fft-comp}, and show that azimuthal wavenumbers in Cartesian runs saturate at levels that are six orders of magnitude weaker compared to polar runs.

Furthermore, we interpolated our fiducial polar run with a fully developed spiral onto a Cartesian grid and noticed that the spiral features of the viscous instability completely vanished within a span of 2--4 orbits after continuation. This suggests that even though numerical diffusion causes an effect akin to physical viscosity, the properties of numerical and physical viscosity are not the same. Both physical and numerical viscosity lead to a viscous spreading of the initial ring distribution, whereas a physical viscosity is necessary for the development of the spiral instability. The same effect cannot be replicated by numerical diffusion alone.
\subsection{Effect of characteristic limiting on numerical viscosity}

\begin{figure}[h]
	\includegraphics[width=\columnwidth]{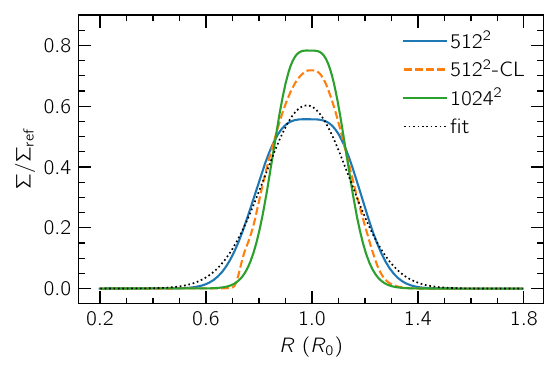}
	\caption{Inviscid Cartesian $512^2$ simulation utilizing characteristic limiting (dashed line) compared to out standard inviscid models at resolution $512^2$ and $1024^2$ (solid lines), after 150 orbits at $R_0$. Characteristic limiting reduces numerical viscosity roughly by a factor of 2 at $512^2$ resolution. It also better maintains local features of the ring, especially the density peak which is rather flattened by numerical viscosity in the compared models. A dotted black line shows the fit using the analytical ring profile for the $512^2$ simulation.}
	\label{fig:1Dcart-512-chr-lim}
\end{figure}

A commonly used method to reduce diffusive numerical effects is characteristic limiting \citep[henceforth ``CL'', implemented by][]{mignone-etal-2012}, which performs spatial reconstruction on characteristic variables instead of the primitive variables in the system. We assessed the effect of this method by comparing to our inviscid $512\times512$ Cartesian simulation with otherwise identical parameters. The low resolution was chosen to highlight the substantial difference while using CL. As shown in \figref{fig:1Dcart-512-chr-lim}, the simulation with CL is less diffusive, evident by the lack of a flattened peak and an overall less-diffused ring. On average, the model with CL had roughly half the numerical viscosity as the standard inviscid $512\times512$ model.

\section{Conclusions}\label{sec:conclusions}

We revisited the viscous ring problem in \citetalias{speith-2003} using the finite-volume codes \texttt{PLUTO} and \texttt{Athena++} as well as the finite-difference code \texttt{FARGO}, in an attempt to measure the numerical viscosity of hydrodynamical codes.

We first reproduced the viscous ring spreading in one dimension and the spiral instability in two dimensions on a polar grid. We analyzed the growth of the instability and showed that \citetalias{speith-2003} had insufficient spatial resolution to properly resolve the instability growth.

We then evolved the viscous ring spreading in two dimensions on a Cartesian grid, applying the same viscosity $\nu$ as used in the polar grid run. The evolution matched the analytical solution to good accuracy but failed to develop the spiral instability even at higher resolutions. We attributed the absence of spiral development to the lack of angular momentum conservation and subsequently high numerical diffusion inherent to Cartesian grids. Our findings suggest that a high physical viscosity is not the only ingredient to developing the spiral instability, but also a numerical setup with very good angular momentum conservation and low numerical diffusivity.

The viscous spreading ring was then used to measure the numerical viscosity in Cartesian grids. This was done by evolving the viscous ring for an inviscid setup and fitting the simulated ring evolution with the analytical solution in \eqref{eq:viscous-ring} as a function of time. As the ring evolution depends only on the viscosity $\nu$, this method allowed us to extract the viscosity over which the system evolved, which for an inviscid system is the numerical viscosity. We then showed a scaling relation between numerical viscosity and resolution that, for second-order methods, corresponds to $\nunumcode\approx 0.63\,\Delta x^2$. Translating our results to a Shakura--Sunyaev $\alpha$ parameter, we found a relation $\alpha\approx 2\times10^{-4}/\sqrt{\bar{R}}$ for our fiducial model with $\nunumcode\approx\ 4\times10^{-7}$ and a realistic aspect ratio $h=0.05$.

We highlight that our models all utilize second-order accurate schemes in both space and time. Even though the effects of numerical viscosity can be mitigated to some degree by using higher order spatial reconstruction and time marching algorithms, further study needs to be done on that matter.

Our results show the existence of moderate diffusive effects in Cartesian grids and quantify the resulting numerical viscosity for standard numerical parameters and different grid resolutions. We also lay out a method that can be used to quantify the numerical viscosity to good accuracy. This information can be utilized to make informed decisions on how to measure and minimize numerical diffusion in hydrodynamics simulations of accretion disks, and is especially useful in the context of low-viscosity or even inviscid configurations of circumbinary or protoplanetary disks on Cartesian grids, as well as for inherently Cartesian codes.

\begin{acknowledgements}
We dedicate this paper to the memory of our dear friend and mentor Willy Kley, 
with whom we developed the idea for this paper. We thank him for his 
fundamental work, counseling, and kindness. AZ and RPN are supported by STFC grant ST/P000592/1, and RPN is supported by the Leverhulme Trust through grant RPG-2018-418. GAT is supported by an STFC PhD studentship.  JJ is co-funded by the European Union (ERC, EPOCH-OF-TAURUS, 101043302). Views and opinions expressed are however those of the author(s) only and do not necessarily reflect those of the European Union or the European Research Council. Neither the European Union nor the granting authority can be held responsible for them.
The authors acknowledge support by the High Performance and
Cloud Computing Group at the Zentrum f\"ur Datenverarbeitung of the University
of T\"ubingen, the state of Baden-W\"urttemberg through bwHPC and the German
Research Foundation (DFG) through grant no INST 37/935-1 FUGG.
This research utilised Queen Mary's Apocrita HPC facility, supported by QMUL Research-IT (http://doi.org/10.5281/zenodo.438045).
This work was performed using the DiRAC Data Intensive service at Leicester, operated by the University of Leicester IT Services, which forms part of the STFC DiRAC HPC Facility (www.dirac.ac.uk). The equipment was funded by BEIS capital funding via STFC capital grants ST/K000373/1 and ST/R002363/1 and STFC DiRAC Operations grant ST/R001014/1. DiRAC is part of the National e-Infrastructure.
The plots in this paper were prepared using the Python library matplotlib 
\citep{hunter-2007}.
\end{acknowledgements}

\bibliography{visc-ring}
\bibliographystyle{aa}

\clearpage
\begin{appendix}
\section{Effect of pressure scale height on the viscous ring spreading and spiral instability}
\label{apdx:aspect_ratio}

\begin{figure}[h]
	\includegraphics[width=\columnwidth]{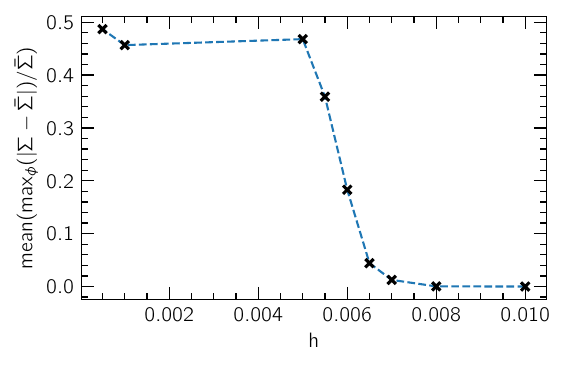}
	\caption{Time average of the maximal spiral amplitude at $R=0.5\,R_0$ 
			between $\tau=0.2$--0.3 for different values of aspect ratio $h$. 
			The spiral instability is strongly damped for $h>0.006$ while spiral 
			amplitudes converge for $h<0.005$.}
	\label{fig:aspect-ratio}
\end{figure}

The analytical solution for the viscous ring spreading and the the stability analysis of the spiral instability studied in \citetalias{speith-2003} assumes a disk without pressure. For a locally isothermal system, this would imply an aspect ratio of $h = 0$ but the \texttt{PLUTO} solver becomes numerically unstable under this condition. To ensure that the pressure in our simulations is small enough to properly simulate the spiral instability we compared the maximal spiral amplitude time averaged over $\tau=0.2$-0.3 for different values of aspect ratio $h$ in \figref{fig:aspect-ratio}.

For aspect ratios $h>0.006$, the spiral instability is strongly damped and the ring structure deviates substantially from the analytical solution. For aspect ratios $h<0.006$, the instability becomes stronger, and is fully saturated at our fiducial aspect ratio of $h=0.005$. We also tested and confirmed (not shown) that the numerical viscosity measured in our Cartesian runs is fully converged for $h=0.005$. Nevertheless, our fiducial aspect ratio is small enough for the gas to be considered pressureless, and we found that the noise we observed in the amplitude of the instability (see e.g., \figref{fig:growth-rates}) fades away for even lower aspect ratios and the time evolution becomes smoother like it is in our \texttt{FARGO} simulations (see \figref{fig:fargo_growth}).

\FloatBarrier
\section{Comparison of codes and the spiral flip in SK03}
\label{sec:spiral_flip}
\begin{figure}[h]
	\includegraphics[width=\columnwidth]{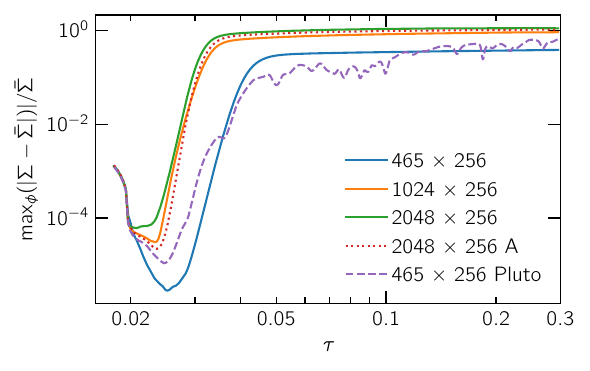}
	\caption{Time evolution of the normalized maximal deviation from the initial
		surface density distribution for different \texttt{FARGO} simulations
		compared to the fiducial \texttt{PLUTO} simulation.}
	\label{fig:fargo_growth}
\end{figure}

To test the robustness of our results on the spiral instability we performed a similar suite of simulations using the \texttt{FARGO} code. \figureref{fig:fargo_growth} shows the growth of the spiral instability for different resolutions. We see that a resolution of $465\times256$ is sufficiently converged, with full convergence at $1024\times256$ and above. Even though \texttt{PLUTO} achieved convergence at a lower resolution of $465\times256$, \texttt{FARGO} shows slight improvements in the growth onset and amplitudes for higher resolutions. Regardless, we can conclude that the results are consistent between both codes. We note that just like \texttt{PLUTO}, the spiral flip vanishes before it reaches the ring in \texttt{FARGO} simulations for resolutions with more that 465 radial cells.

\begin{figure*}[h]
	\centering
	\includegraphics[width=\textwidth]{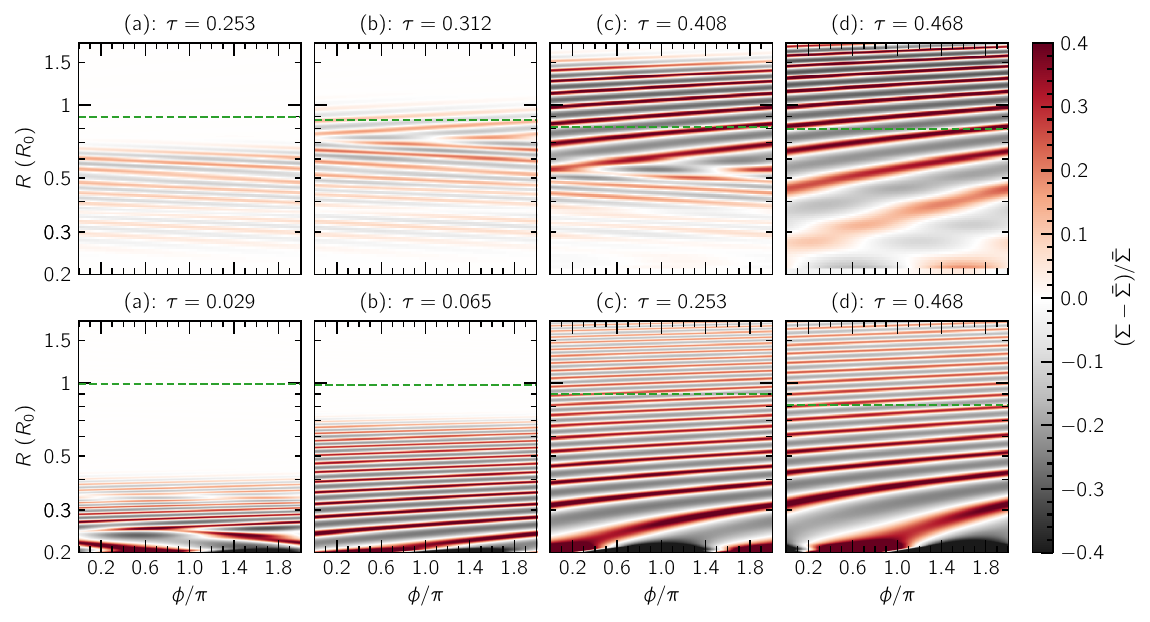}
	\caption{Surface density deviation heatmaps similar to Fig.~\ref{fig:2Dpol-std-evo}, for \texttt{FARGO} simulations at a resolution of $256\times256$ on an arithmetic grid (top) and at $456\times256$ on a logarithmic grid (bottom). Top: this setup reproduces the results of \citetalias{speith-2003} (see Fig.~6 therein). The flip from a leading to trailing spiral is visible in the middle two panels, before it disappears into a single leading spiral. Bottom: this setup is equivalent to our fiducial \texttt{PLUTO} model, and behaves very similar to our results in Fig.~\ref{fig:2Dpol-std-evo}. Here, spirals develop earlier and the initial flip is short-lived and closer to the inner boundary.}
	\label{fig:willy_compare_sample}
\end{figure*}

To further understand the spiral flip results shown in 
\citetalias{speith-2003} we replicated their setup in \texttt{FARGO}, with a 
$256\times256$ arithmetically spaced polar grid. Our simulation reproduces 
their Fig.\,6 very well, but is run for longer. The spiral 
flip pattern emerges after $\tau=0.04$ and gradually morphs into a 
single leading spiral for $\tau>0.43$ (see \figref{fig:willy_compare_sample}).
We conducted additional simulations with higher resolutions and
also on logarithmic grids. Each time the resolution was increased on an arithmetic grid,
the flip occurred further inward and turns into a leading spiral earlier into the simulation.
On a logarithmic grid, the inner region is better resolved and the flip always 
vanishes early into the simulation and never reaches the spreading ring, see 
\figref{fig:willy_compare_sample}.

\begin{figure}[h]
	\includegraphics[width=\columnwidth]{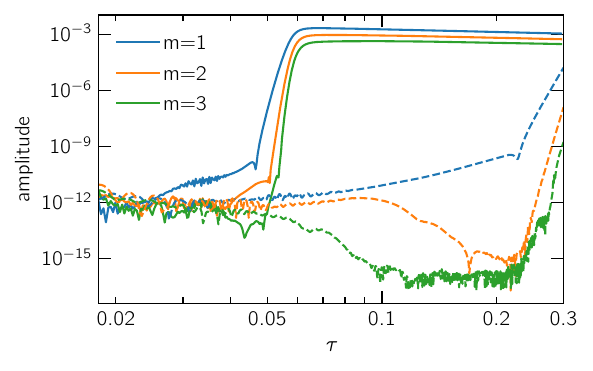}
	\caption{Time evolution of the first three Fourier azimuthal modes at 
	$R=R_0$. Solid lines represent a high-resolution ($2048\times256$) simulation using a logarithmic grid, and the dashed lines represent results from an 
	identical setup as in \citetalias{speith-2003}, with a $256\times256$ arithmetic grid.}
	\label{fig:fargo_comparewillygrowth}
\end{figure}

A comparison of the growth rates between the \citetalias{speith-2003}
setup and a high resolution \texttt{FARGO} run is shown in 
\figref{fig:fargo_comparewillygrowth}. As the grid is the only difference 
between the simulations, the flip seems to be a numerical effect that is 
exacerbated when using an arithmetic grid, which has a lower resolution at 
the inner boundary.
The constant smoothing length used for the SPH code in \citetalias{speith-2003} has the same resolution 
effect as an arithmetic grid, which explains why their simulations are in 
agreement. We conclude that the flipped spiral structure with a 
trailing spiral observed at the inner domain in	\citetalias{speith-2003} is 
merely a numerical effect.
	 
\FloatBarrier
\section{Numerical diffusion as a consequence of advection}
\label{apdx:num-visc}
We consider the 1D advection equation of a quantity $q$ for a fluid moving at constant velocity $u>0$
\begin{equation}
	\label{eq:advection}
	\DP{q}{t} + u\DP{q}{x} = 0.
\end{equation}

We can discretize this equation to a first-order upwind scheme \citep[e.g.,][]{courant1952} where $q^\prime = q(t+\Delta t)$ on a grid where $i$ is the cell index such that $q_{i-1} = q(x_i - \Delta x)$
\begin{equation}
	\label{eq:upwind}
	\frac{q_i^\prime - q_i}{\Delta t} + u\frac{q_i - q_{i-1}}{\Delta x} = 0.
\end{equation}
By Taylor-expanding $q_i^\prime$ and $q_{i-1}$ to second order in time and space respectively, and substituting a wave-like solution $q_{tt} = uq_{xx}$, we arrive at a modified upwind equation:
\begin{equation}
	\label{eq:diffusive-upwind}
	\DP{q}{t} + u\DP{q}{x} = D\frac{\partial^2q}{\partial x^2},\qquad D = \frac{u\Delta x}{2} (1-C),\quad C\equiv u\frac{\Delta t}{\Delta{x}}
\end{equation}
which corresponds to an advection--diffusion equation with diffusion coefficient $D$. This term, while not physically motivated, allows the upwind scheme to remain stable for Courant numbers $0<C<1$.

This approach does not exactly relate to our results since the gas velocity is not constant and we further use second-order schemes with \texttt{PLUTO}. Nevertheless, we can expect that our numerical scheme will give rise to a similarly-motivated numerical diffusion term such that \eqref{eq:hydro} effectively becomes
\begin{equation}
	\DP{(\Sigma\vel)}{t} + \nabla\cdot\left(\Sigma\vel\otimes\vel\right) = -\nabla P - \Sigma\nabla\Phi_\star + \nabla\cdot(\tensorGR{\upsigma}+\tensorGR{\upsigma}_\mathrm{num}),
\end{equation}
with $\tensorGR{\upsigma}\propto \nu$ and
$\tensorGR{\upsigma}_\mathrm{num}\propto \nu_\text{num}$. Given that our
inviscid Cartesian models behave similarly to viscous models with viscosity
$\nu_\text{num}$ as far as ring spreading is concerned, it is therefore
unsurprising that a ring in a viscous model with $\nu\sim \nu_\text{num}$ should spread as if the total viscosity were $\nu_\text{tot} \approx \nu + \nu_\text{num}$.

We also note that, for a given Courant number $C$, the diffusion coefficient in 
\eqref{eq:diffusive-upwind} is proportional to $\Delta x$ for the first-order 
method we considered, and expect a $\Delta x^2$ scaling for a second-order 
method in 1D. Given that the viscous ring problem evolves on both the $x$ and 
$y$ directions in 2D, doubling the resolution would require increasing the 
number of cells in both directions by a factor of 2 (thus increasing grid cell 
count by a factor of 4). In doing so, we would expect a scaling $\nunum\propto 
\Delta x^2$, which is verified in \figref{fig:numerical-viscosity}.

Finally, we found that doubling the resolution in only one direction (such as $N_x\times N_y = 512\times1024$) results in a numerical viscosity estimate that is much closer to the value expected for the low-resolution direction (in this case, $N_x$) than to an estimate dictated by $\Delta x\times \Delta y$. This rules out that the numerical diffusion we observe depends linearly on $\Delta x$ and $\Delta y$ and suggests that it more accurately depends on $\min(\Delta x, \Delta y)^2$, highlighting the second-order accuracy of our solver.
\FloatBarrier
\section{Measuring numerical viscosity}
\label{apdx:measuring-num-visc}
To measure the numerical viscosity in our inviscid simulations we utilize the relation between the viscous time scale $\tau$ for the ring spreading and the orbital time t at $R_0$ given by $\tau = 12 \nu t / R^2_0$. We note that for a given viscosity $\nu$, this equation represents the slope--intercept form of an equation for a line with slope $m = 12\nu/R_0^2$ and intercept zero.

\begin{figure}[h]
	\includegraphics[width=\columnwidth]{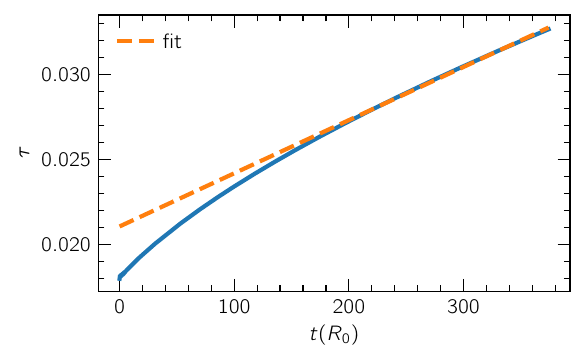}
	\caption{The $\tau(t)$ curve for our inviscid 1024 model. The dashed line 
			shows the fit over the last 180 time units. The slope of this line 
			is used to extract the numerical viscosity.}
	\label{fig:tau-fit}
\end{figure}

To do so, first we fit the analytical solution to a slice of the two-dimensional data at time $t$ to compute $\tau$. Repeating this procedure for all snapshots generates a curve for $\tau(t)$ shown in \figref{fig:tau-fit}. We then fit a straight line to the $\tau(t)$ curve for the last $\sim$ 180 orbits. The slope of this straight line fit gives the numerical viscosity using the above relation for the slope $m$.
\FloatBarrier
\section{Inviscid Cartesian simulations with \texttt{Athena++}}
\label{apdx:athena}

For the suite of Cartesian models using \texttt{Athena++} we implement an 
equivalent setup to the \texttt{PLUTO} runs, as described in 
\secref{sec:model}. A locally isothermal disk is achieved by using an adiabatic 
equation of state combined with a thermal relaxation timescale 
$\tau_\textrm{cool} = 0.01\,\OmegaK^{-1}$. The viscous ring sits on top of a 
constant density background with $\Sigma_\textrm{bgr} = 
10^{-7}\,\Sigma_\textrm{ref}$, however the density floor of the domain was set 
to $\Sigma_\textrm{floor} = 10^{-15}\,\Sigma_\textrm{ref}$ as interactions with 
the density floor created numerical instabilities.

Our results are listed in Table~\ref{tab:numerical_viscosities} and 
shown in Fig.~\ref{fig:numerical-viscosity}. Both codes agree very well 
for $N_\textrm{cells}\geq1024^2$, which corresponds to our fiducial resolution. 
The expected behavior of $\nunum \propto \Delta x^2$ is also found with 
\texttt{Athena++}.

\FloatBarrier
\end{appendix}
\end{document}